\documentclass[aps,prb,twocolumn,groupedaddress]{revtex4-2}

\usepackage[T1]{fontenc}
\usepackage{amsmath}
\usepackage{amssymb}
\usepackage{bbm}
\usepackage{xspace}
\usepackage{color}
\usepackage{graphicx}
\usepackage{ulem} % to strike out text
\usepackage{placeins}
\usepackage{bm}
\usepackage{siunitx}
\usepackage{nccmath}
\usepackage{mathtools}   % loads »amsmath«
\usepackage{empheq}
\usepackage{enumerate}

\usepackage[colorlinks=true,linkcolor=magenta,citecolor=blue,urlcolor=blue]{hyperref}

\usepackage{times}
\usepackage[version=4]{mhchem}

\graphicspath{{.}{Bilder/}}

\begin{document}

\title{Andreev reflection of massive pseudospin-1 fermions}
\author{W.~Zeng$^1$}
\author{R.~Shen$^{1,2}$}
\email[E-mail: ]{shen@nju.edu.cn}
\affiliation{$^1$National Laboratory of Solid State Microstructures and School of Physics, Nanjing University, Nanjing 210093, China\\
$^2$Collaborative Innovation Center of Advanced Microstructures, Nanjing University, Nanjing, 210093, China}

\date{\today}

\begin{abstract}{}
We theoretically investigate the Andreev reflection of the pseudospin-1 Dirac fermions with either the $\pm U$-type or the $S_z$-type mass term. For the $\pm U$-type fermions, it is found that the Andreev reflection probability at the oblique incidence can be even larger than that at the normal incidence. For the retro-reflection, such an oblique enhancement occurs in the $n$-doped $+U$-type ($p$-doped $-U$-type) massive fermion systems. While for the specular reflection, the enhancement occurs in the $n$-doped $-U$-type ($p$-doped $+U$-type) systems. For the $S_z$-type massive fermions, a super Andreev reflection with all-angle unit efficiency is predicted in an undoped junction with the incident energy equal to the superconducting gap. 
\end{abstract}

\maketitle
\section{Introduction}\label{intro}
Dirac materials become one of the hot spots in condensed matter physics after the discovery of graphene\cite{PhysRevLett.97.067007,PhysRevB.98.075301,PhysRevB.83.235403,PhysRevLett.108.106603}, in which the carbon atoms are assembled into a two-dimensional honeycomb lattice. The conduction and the valence bands in graphene meet at six Dirac points, where the low energy excitations are the massless pseudospin-$1/2$ Dirac fermions\cite{RevModPhys.80.1337,PhysRevB.78.193406,PhysRevB.101.045407,PhysRevB.75.195322,PhysRevB.104.075436}. Apart from graphene, materials hosting the pseudospin-1 fermions have also attracted considerable attention, such as the two-dimensional $\mathcal{T}_3$ lattice\cite{PhysRevA.80.063603,PhysRevA.83.023609,PhysRevB.81.041410}, Lieb lattices\cite{PhysRevB.88.161413,doi:10.1063/1.3526724}, and breathing lattices\cite{essafi2017flat}. The low energy excitations of these pseudospin-1 fermions are featured by a flat band cutting through two linearly dispersing branches at the Dirac points. The flat band related physical phenomena have been widely reported, such as the super-Klein tunneling\cite{PhysRevB.96.024304,PhysRevB.84.115136,PhysRevB.93.035422,PhysRevB.103.195439,MANDAL2020126666}, the Landau-Zener Bloch oscillations\cite{PhysRevLett.116.245301}, the ﬂat-band induced conductivity\cite{PhysRevB.88.161413,PhysRevB.91.041102,PhysRevB.92.155116}, and the unconventional Anderson localization\cite{PhysRevB.82.104209}.

\begin{figure}[tb]
\centerline{\includegraphics[width=1\linewidth]{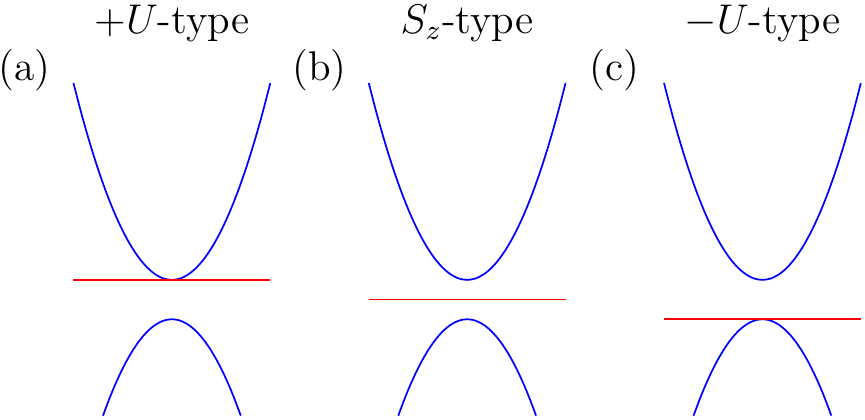}}
\caption{\label{schematic}
Schematic diagram of the energy dispersion for the massive pseudospin-1 fermions. The flat band is indicated by the red lines.
}
\end{figure}

An additional mass term can be introduced in the pseudospin-1 Dirac systems, which opens a band gap with a flat band in it. There are three types of mass terms usually studied in the literature\cite{PhysRevB.96.024304,PhysRevB.103.195442}, namely the $\pm U$-type and the $S_z$-type, respectively. For the $\pm U$-type massive pseudospin-1 fermions, the flat band is located at the top (bottom) of the band gap, as shown in Figs.\ \ref{schematic}(a) and \ref{schematic}(c), respectively. For the $S_z$-type one, the flat band is located at the center of the gap, as shown in Fig.\ \ref{schematic}(b). Several methods have been proposed to produce the massive pseudospin-1 fermions in the cold atom or the photonic lattices, such as tuning the on-site energy difference or the dimerization interaction between the corner and the edge-center sites in the Lieb lattice and the $\mathcal{T}_3$ lattice\cite{PhysRevB.81.041410,PhysRevB.82.085310,PhysRevA.83.063601,PhysRevB.86.195129,PhysRevB.82.075104,PhysRevResearch.2.013062,romhanyi2015hall}. There are also some realistic two-dimensional materials which host the massive pseudospin-1 fermions, such as the $\ce{SrTiO3}/\ce{SrIrO3}/\ce{SrTiO3}$ trilayer heterostructure grown along the ($111$) direction\cite{PhysRevB.84.241103} and the two-dimensional $sp^2$ carbon-conjugated covalent-organic framework ($sp^2$-c-COF)\cite{jiang2019lieb}. 

Very recently, Feng \textit{et al.}\cite{PhysRevB.101.235417} and Zhou\cite{PhysRevB.104.125441} have studied the Andreev reflection in the massless pseudospin-1 system. Both of them predicted the super Andreev reflection effect, where the Andreev reflection with the unit probability appears independent of the incident angles. It is natural to ask how the mass terms affect the Andreev reflection of the pseudospin-1 fermions and whether the super Andreev reflection is preserved or not. Motivated by this, we report a study on the Andreev reflection of the massive pseudospin-1 fermions. It is shown that, distinct from the massless Dirac systems\cite{PhysRevLett.97.067007,PhysRevB.101.235417,PhysRevB.104.125441}, the Andreev reflection of the $\pm U$-type massive fermions exhibits an unusual enhancement with increasing the incident angle. For the $+U$-type fermions, such an oblique enhancement appears in the retro-reflection in an $n$-doped system and in the specular reflection in a $p$-doped system. For the $-U$-type fermions, the oblique enhancement appears in the retro-reflection in a $p$-doped system and in the specular reflection in an $n$-doped system. In the undoped systems, the super Andreev reflection still survives for the $S_z$-type fermions with the incident energy at the superconducting gap. For the $\pm U$-type fermions, the super Andreev reflection is attenuated, which means that one can achieve a mass dependent Andreev reflection probability less than one, independent of the incident angles. 

The rest of this paper is organized as follows. The model is explained in Sec.\ \ref{model}. The numerical results and discussions are presented in Sec.\ \ref{results}. Finally, we conclude in Sec.\ \ref{conclusions}.

\section{Model}\label{model}
We consider the normal metal/superconductor junction based on the $\mathcal{T}_3$ model. The junction lies in the $x$-$y$ plane with the current along the $x$ direction. The superconductivity in the right side of the junction is induced by the proximity effect of a superconducting electrode covering the region $x>0$. The single particle Hamiltonian reads\cite{PhysRevB.104.125441,PhysRevB.96.024304} 
\begin{align}
\mathcal{H}_{\pm}=\hbar v_F(\pm S_xk_x+S_yk_y)+m \Gamma+V(x),\label{equsingle}
\end{align}
where the subscripts $\pm$ denote two valleys, $v_{F}$ is the Fermi velocity, $k_{x,y}$ are the wave vectors and $S_{x,y}$ are the spin-1 matrices given by
\begin{align}
S_x=\frac{1}{\sqrt{2}}\begin{pmatrix}
0 &1 &0\\ 
1 &0 &1 \\ 
0 &1 &0 
\end{pmatrix},\quad S_y=\frac{1}{\sqrt{2}}\begin{pmatrix}
0 &-i &0\\ 
i &0 &-i \\ 
0 &i &0 
\end{pmatrix}.
\end{align}
The electrostatic potential $V(x)$ is zero in the normal region and $V_{0}$ in the superconducting region, which can be adjusted by doping or by a gate voltage. The mass term is described by the mass amplitude $m$ and the $\Gamma$ matrix, which is given by\cite{PhysRevB.96.024304,PhysRevB.103.195442}
\begin{align}
\Gamma=\pm\begin{pmatrix}
1 &0 &0 \\ 
0 &-1 &0\\ 
0 &0 &1
\end{pmatrix}\quad \text{or}\quad \Gamma=\begin{pmatrix}
1 &0 &0 \\ 
0 &0 &0\\ 
0 &0 &-1
\end{pmatrix},
\end{align}
representing either the $\pm U$-type or the $S_z$-type massive fermions, respectively. 

The energy dispersion for the $+U$-type fermions in the normal region is given by 
\begin{align}
\varepsilon_s=s\sqrt{(\hbar v_F k)^2+m^2},\quad \varepsilon_\mathrm{flat}=m,\label{u}
\end{align}
where $s=\pm 1$ denotes the conduction and the valence bands, respectively. The flat band is located at the bottom of the conduction band. By replacing $m$ with $-m$ in Eq.\ (\ref{u}), one obtains the energy dispersion for the $-U$-type fermions, where the flat band is located at the top of the valence band. The energy dispersion for the $S_z$-type fermions is given by 
\begin{align}
\varepsilon_s=s\sqrt{(\hbar v_F k)^2+m^2},\quad \varepsilon_\mathrm{flat}=0,\label{sz}
\end{align}
where the flat band is located at the center of the band gap.

The normal metal/superconductor junction is described by the valley decoupled Bogoliubov-de Gennes (BdG) equation due to the time reversal symmetry\cite{de2018superconductivity}
\begin{align}
\begin{pmatrix}
\mathcal{H}_{\pm}-\mu & \Delta(x)\\ 
\Delta^\dagger(x) & \mu-\mathcal{H}_{\pm}
\end{pmatrix}\begin{pmatrix}
\psi_e\\ 
\psi_h
\end{pmatrix}=\varepsilon\begin{pmatrix}
\psi_e\\ 
\psi_h
\end{pmatrix},\label{equbdg}
\end{align}
where $\varepsilon$ is the excitation energy measured from the Fermi level $\mu$, $\psi_{e(h)}$ is the electron (hole) component of the quasiparticle wave function, respectively. Due to the valley degeneracy, it is sufficient to consider the set with $\mathcal{H}_+$ in the BdG equation. The $s$-wave superconducting pair potential $\Delta(x)$ is introduced by the proximity effect\cite{VOLKOV1995261,efetov2016specular}, which is assumed zero in the normal region and a real constant $\Delta$ in the superconducting region. Since the two spin channels are also decoupled in the BdG equation, the true spin indices are dropped in Eq.\ (\ref{equbdg}).

By solving the BdG equation (\ref{equbdg}), the electron state and the hole state in the normal region are given by ($\hbar v_F=1$)
\begin{align}
\psi_{e}^\pm=
&\begin{pmatrix}
\frac{\pm s_e k_e-ik_y}{\varepsilon+\mu-\varsigma} &\sqrt{2}  & \frac{\pm s_e k_e+ik_y}{\varepsilon+\mu-\varrho}&0&0&0
\end{pmatrix}^T\nonumber \\
&\times e^{\pm is_ek_ex+ik_yy},\label{fe}\\
\psi_{h}^\pm=
&\begin{pmatrix}
0&0&0&\frac{\pm s_h k_h-ik_y}{\varepsilon-\mu+\varsigma} &-\sqrt{2}  & \frac{\pm s_h k_h+ik_y}{\varepsilon-\mu+\varrho}
\end{pmatrix}^T\nonumber\\ 
&\times e^{\pm is_{h}k_hx+ik_yy},\label{fh}
\end{align}
where the particle states propagating along the $\pm x$ directions are labeled by the superscript $\pm$, respectively. The sign index $s_{e(h)}$ distinguishes the electron and hole states in the conduction band from those in the valence band. For the electron state and the hole state in the conduction band, we have $s_{e}=+1$ and $s_{h}=-1$, respectively. For those in the valence band, we have $s_{e}=-1$ and $s_{h}=+1$, respectively. The longitudinal wave vectors for the electron and the hole states are $k_{e(h)}=\sqrt{(\varepsilon\pm\mu)^2-m^2-k_y^2}$. The parameters $(\varsigma,\varrho)$ are defined as $(\pm m,\pm m)$ for the $\pm U$-type fermions and $(m,-m)$  for the $S_z$-type fermions, respectively. 

\begin{figure*}[tb]
\centerline{\includegraphics[width=1\linewidth]{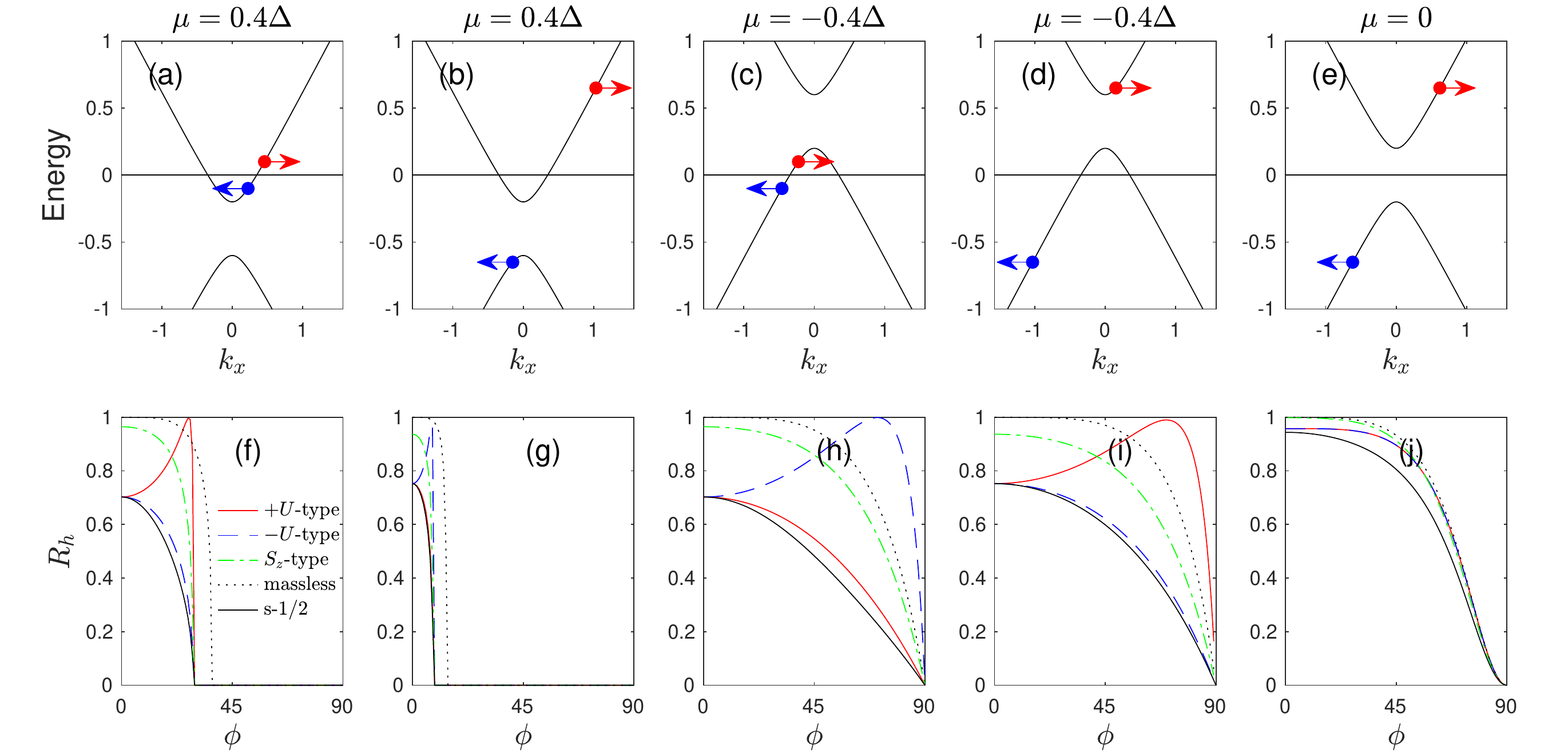}}
\caption{\label{f2}
(a)-(e) Andreev reflection schematics. The red and blue bullets with arrows indicate the incident electron state and the Andreev reflected hole state in the energy band in the normal region, respectively, where the flat band is omitted and the location of the Fermi level is represented by the horizontal line at center. (f)-(j) Andreev reflection probabilities as a function of the incident angle corresponding to the processes in (a)-(e), respectively. The incident energy is $\varepsilon=0.1\Delta$ for the cases of (a) and (c) and $\varepsilon =0.65\Delta$ for the cases (b), (d), and (e), respectively. The mass amplitude is $m=0.2\Delta$ in all cases.
}
\end{figure*}

Under the heavily doping condition, the scattering state in the superconducting region is given by
\begin{align}
\psi_s^\pm=&\begin{pmatrix}
1 & \pm\sqrt{2} &1  &\frac{\varepsilon+\mu_s\mp k^s_\pm}{\Delta}  &\pm\sqrt{2}\frac{\varepsilon+\mu_s\mp k^s_\pm}{\Delta}  & \frac{\varepsilon+\mu_s\mp k^s_\pm}{\Delta}
\end{pmatrix}^T\nonumber\\
&\times e^{ik^s_\pm x+ik_yy},\label{fs}
\end{align}
where the superscript $\pm$ denotes the electronlike and the holelike quasiparticle states, respectively. The longitudinal wave vector in the superconducting region is given by $k^s_\pm=\pm\mu_s+\sqrt{\varepsilon^2-\Delta^2}$ with $\mu_s=\mu+V_0$.

Considering an electron coming from the left, the wave functions in the normal and superconducting regions are given by
\begin{align}
&\psi_N=\psi_e^++r_h\psi_h^-+r_e\psi_e^-,\label{wn}\\
&\psi_S=t_+\psi_s^++t_-\psi_s^-,\label{ws}
\end{align}
where $t_{\pm}$ is the transmission amplitude for the electron-like (hole-like) quasiparticles and $r_{h(e)}$ is the Andreev (normal) reflection amplitude, respectively. The boundary conditions connecting $\psi_N$ and $\psi_S$ can be obtained by integrating Eq.\ (\ref{equbdg}) over a small interval around $x=0$. Given a general six-component wave function $(\psi_A^e,\psi_B^e,\psi_C^e,\psi_A^h,\psi_B^h,\psi_C^h)^T$, the boundary conditions are obtained as \cite{PhysRevB.84.115136}
\begin{align}
\psi_B^{e(h)}(0^-)&=\psi_B^{e(h)}(0^+),\label{boundary1}\\
\psi_A^{e(h)}(0^-)+\psi_C^{e(h)}(0^-)&=\psi_A^{e(h)}(0^+)+\psi_C^{e(h)}(0^+).\label{boundary2}
\end{align}

By substituting Eqs.\ (\ref{wn}) and (\ref{ws}) into Eqs.\ (\ref{boundary1}) and (\ref{boundary2}), the reflection amplitudes are obtained as
\begin{align}
&r_h=\frac{\zeta^e_+-\zeta^e_-}{\left(\zeta^e_-+1 \right )\left(\zeta^h_-+1 \right )\zeta^s_--\left(\zeta^e_--1 \right )\left(\zeta^h_--1 \right )\zeta^s_+},\label{pre}\\
&r_e=\frac{\left(\zeta^e_+-1 \right )\left(\zeta^h_--1 \right )\zeta^s_+-\left(\zeta^e_++1 \right )\left(\zeta^h_-+1 \right )\zeta^s_-}{\left(\zeta^e_-+1 \right )\left(\zeta^h_-+1 \right )\zeta^s_--\left(\zeta^e_--1 \right )\left(\zeta^h_--1 \right )\zeta^s_+}\label{prh},
\end{align}
where 
\begin{align}
&\zeta_\pm^e=\frac{1}{2}\left(\frac{\pm s_ek_e-ik_y}{\varepsilon+\mu-\varsigma}+\frac{\pm s_ek_e+ik_y}{\varepsilon+\mu-\varrho}\right),\label{z1}\\
&\zeta_\pm^h=\frac{1}{2}\left(\frac{\pm s_hk_h-ik_y}{\varepsilon-\mu+\varsigma}+\frac{\pm s_hk_h+ik_y}{\varepsilon-\mu+\varrho}\right),\label{z2}\\
&\zeta_\pm^s=\frac{1}{2}\left(\frac{\varepsilon\pm\sqrt{\varepsilon^{2}-\Delta^{2}}}{\Delta}\right).\label{z3}
\end{align}
The normal and the Andreev reflection probabilities can be obtained by 
\begin{align}
&R_e=\left|\frac{\langle\psi_e^-|v_x|\psi_e^-\rangle}{\langle\psi_e^+|v_x|\psi_e^+\rangle}\right||r_e|^2,\label{re}\\
&R_h=\left|\frac{\langle\psi_h^-|v_x|\psi_h^-\rangle}{\langle\psi_e^+|v_x|\psi_e^+\rangle}\right||r_h|^2,\label{rh}
\end{align}
where $v_x=\mathrm{diag}(S_x,-S_x)$ is the velocity operator. With the help of the Blonder-Tinkham-Klapwijk approach\cite{PhysRevB.25.4515}, the differential conductance is obtained as 
\begin{align}
G=G_0\int_0^{\pi/2}\left(1-R_e+R_h\right)\cos\phi\,\mathrm{d}\phi,\label{GG}
\end{align}
where $\phi=\arcsin\big(k_y/\sqrt{(\varepsilon+\mu)^2-m^2}\big)$ is the incident angle, $G_0=4e^2N(\varepsilon)/h$ is the ballistic conductance with $N(\varepsilon)=W\sqrt{(\varepsilon+\mu)^2-m^2}/\pi$ being the number of the transverse modes in the junction of width $W$.

\section{Results}\label{results}
\subsection{Andreev reflection}
The Andreev reflection properties depend on the Fermi level and the incident energy. There are five typical scenarios as depicted in Figs.\ \ref{f2}(a)-\ref{f2}(e), respectively. When the Fermi level is above the gap center ($\mu >0$), the junction is in the $n$-doped region as shown in Figs.\ \ref{f2}(a) and \ref{f2}(b). When the Fermi level is below the gap center ($\mu <0$), the junction is in the $p$-doped region as shown in Figs.\ \ref{f2}(c) and \ref{f2}(d). Depending on the energy of the incident electron, the Andreev reflected hole and the incident electron may either in the same band resulting in a retro-reflection as shown in Figs.\ \ref{f2}(a) and \ref{f2}(c) or in the different bands resulting in a specular reflection as shown in Figs.\ \ref{f2}(b) and \ref{f2}(d). For the undoped case, the Fermi level is just located at the gap center ($\mu =0$) and only the specular reflection can occur, as shown in Fig.\ \ref{f2}(e). 

The Andreev reflection probability can be numerically calculated from Eq.\ (\ref{rh}) and is plotted as a function of the incident angle in Figs.\ \ref{f2}(f)-\ref{f2}(j) corresponding to scenarios \ref{f2}(a)-\ref{f2}(e), respectively. We also present two reference curves in Fig.\ \ref{f2}. One is the Andreev reflection probability of the massless pseudospin-1 fermions which is obtained by taking the limit $m=0$, the other is the results for the massive pseudospin-1/2 fermions which are obtained by a similar approach in Ref. \cite{PhysRevLett.97.067007}.

For the massless pseudospin-1 fermions, there is the perfect Andreev reflection ($R_{h}=1$) at normal incidence as shown by the black dotted lines in Fig.\ \ref{f2}. Since the mass term breaks the pseudospin-momentum locking, such a perfect Andreev reflection is absent for the massive pseudospin-1 fermions.

From Figs.\ \ref{f2}(a) and \ref{f2}(b), one finds that there is a critical angle $\phi_{c}$ for the Andreev reflection in the $n$-doped region. When the incident angle beyonds $\phi_{c}$, the Andreev reflection vanishes. With the help of the expressions of $k_{e(h)}$, the critical angle is obtained as $\phi_c=\arcsin\left[\sqrt{\frac{(\varepsilon-\mu)^2-m^2}{(\varepsilon+\mu)^2-m^2}}\right]$. For the Andreev reflection in the $p$-doped region, the critical angle is extended to $\pi /2$ as shown in Figs.\ \ref{f2}(c) and \ref{f2}(d).

From Figs.\ \ref{f2}(f) and \ref{f2}(g), one can find an unusual enhancement of the Andreev reflection with increasing the incident angle. As shown by the red solid line in Fig.\ \ref{f2}(f), the retro-Andreev reflection of the $+U$-type fermions in an $n$-doped junction increases with the incident angle until the incident angle is close to $\phi_{c}$. For the specular Andreev reflection in the $n$-doped junction, the oblique enhancement occurs with the $-U$-type fermions, as shown by the blue dashed line in Fig.\ \ref{f2}(g).
 
In the $p$-doped junction, similar oblique enhancement is presented. The retro-reflection of the $-U$-type fermions increases with the incident angle until the incident angle is close to $\pi /2$ and the specular reflection of the $+U$-type fermions is also enhanced by the oblique incidence, as shown in Figs.\ \ref{f2}(h) and \ref{f2}(i), respectively.  

In the undoped junction, the Andreev reflection probabilities of the $\pm U$-type fermions are identical due to the particle-hole symmetry, however, the oblique enhancement is absent as shown in Fig.\ \ref{f2}(j). The oblique enhancement of the Andreev reflection is unique to the $\pm U$-type massive pseudospin-1 fermions and absent in the $S_{z}$-type, the massless pseudospin-1 or the pseudospin-1/2 systems.

The oblique enhanced Andreev reflection of the $\pm U$-type fermions can be understood in an analytical way. At a small incident angle, the Andreev reflection probability for the $+U$-type fermions can be expanded as 
\begin{gather}
R_h=R_0+\kappa\phi^2.\label{expansion1}
\end{gather} 
The Andreev reflection probability at normal incidence $R_0$ is given by
\begin{gather}
R_0=\frac{4\mathcal{D_+}\mathcal{D_-}}{\left(1+\mathcal{D_+}\mathcal{D_-}\right)^2\cos^2\beta+\left(\mathcal{D}_++\mathcal{D}_-\right)^2\sin^2\beta},\label{rhnormal}
\end{gather}
where $\mathcal{D}_\pm=\sqrt{\frac{\varepsilon \pm\mu+m}{\varepsilon \pm\mu-m}}$ and $\beta=\arccos(\varepsilon /\Delta)$. The coefficient of $\phi^2$ is given by 
\begin{align}
\kappa=&-\mu m s_e s_h \mathcal{P}
\end{align}
with
\begin{align}
\mathcal{P}=R_0^2\frac{2\varepsilon^{2}\cos^{2}\beta+\sin^{2}\beta\left(\varepsilon^{2}+\mu^{2}-m^{2}\right)}{\left(\mathcal{D}_+/\mathcal{D}_-\right)^2\left(\varepsilon^{2}-\left(\mu-m\right)^2\right)^2\sin\phi_c}.\label{expansion2}
\end{align}
The expression $\varepsilon^{2}+\mu^{2}-m^{2}$ is just $k_{e}^{2}+k_{h}^{2}$ at normal incidence and is always positive. As a result, $\mathcal{P}$ is positive definite for subgap incident energy and the sign of $\kappa$ is only determined by $\mu m s_e s_h$. In the retro-reflection, the electron-hole conversion occurs in the same band with $s_e s_h=-1$ and the oblique enhancement is presented in the $n$-doped region. In the specular reflection, the electron-hole conversion occurs in the different bands with $s_e s_h=+1$ and the oblique enhancement is presented in the $p$-doped region. For the $-U$-type fermions, one can obtain the same expansion with $m$ replaced by $-m$ in Eqs.\ (\ref{expansion1}-\ref{expansion2}). It can be easily found that $R_{0}$ of the $-U$-type fermions is the same as that of the $+U$-type fermions. The condition for the oblique enhancement of the $-U$-type fermions is just opposite to that of the $+U$-type fermions.

\begin{figure}[tb]
\centerline{\includegraphics[width=1\linewidth]{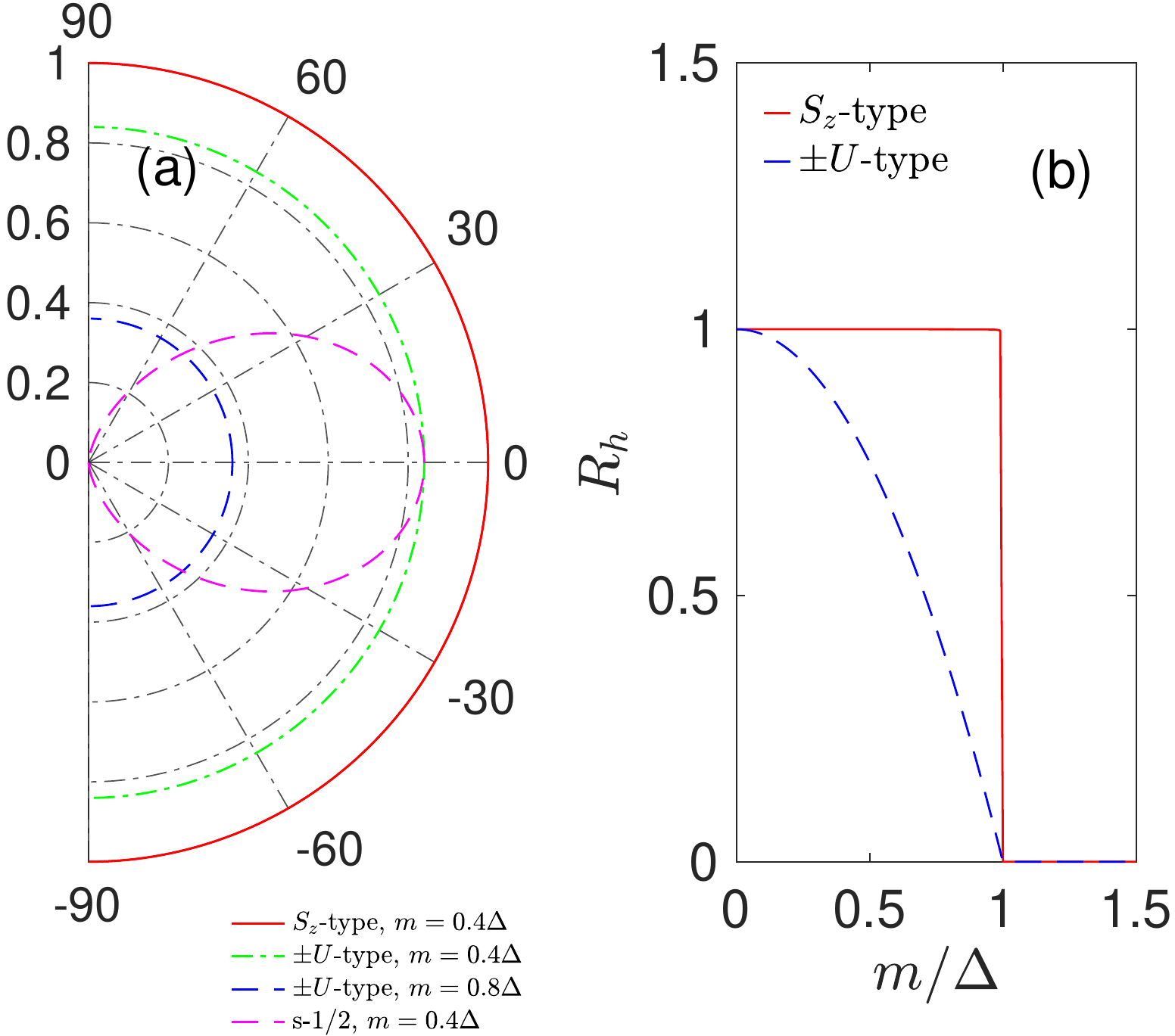}}
\caption{\label{f5}
(a) Polar plot of the Andreev reflection probability with $\mu=0$ and $\varepsilon=\Delta$. (b) Andreev reflection probability as a function of $m$.}
\end{figure}

Next, we discuss the super Andreev reflection in the undoped junction when the incident energy of the electron approaches to the superconducting gap. In Fig.\ \ref{f5}, it is shown that the Andreev reflection probability is independent of the incident angles with $\varepsilon =\Delta$. For the $S_{z}$-type fermions, the Andreev reflection always reaches the unit efficiency, which is called the super Andreev reflection. For the $\pm U$-type fermions, the super Andreev reflection is attenuated. Although the Andreev reflection probability is still independent of the incident angles, it is less than one, depending on the mass amplitude. 

In fact, under the limit of $\mu=0$ and $\varepsilon=\Delta$, Eq.\ (\ref{rh}) can be reduced to
\begin{align}
R_h=1-\mathcal{C}\left(\frac{m}{\Delta} \right )^2 \label{od1},
\end{align}
where $\mathcal{C}=0$ and $1$ for the $S_z$-type and the $\pm U$-type fermions, respectively. Eq.\ (\ref{od1}) is valid only in the region of $m<\Delta$. For $m>\Delta$, there is no subgap Andreev reflection in the undoped junction. The super and the attenuated super Andreev reflections are clearly shown in Eq.\ (\ref{od1}). By comparison, there is no super Andreev reflection in the pseudospin-1/2 system without a flat band, as shown in In Fig.\ \ref{f5}(a).

\begin{figure}[tb]
\centerline{\includegraphics[width=1\linewidth]{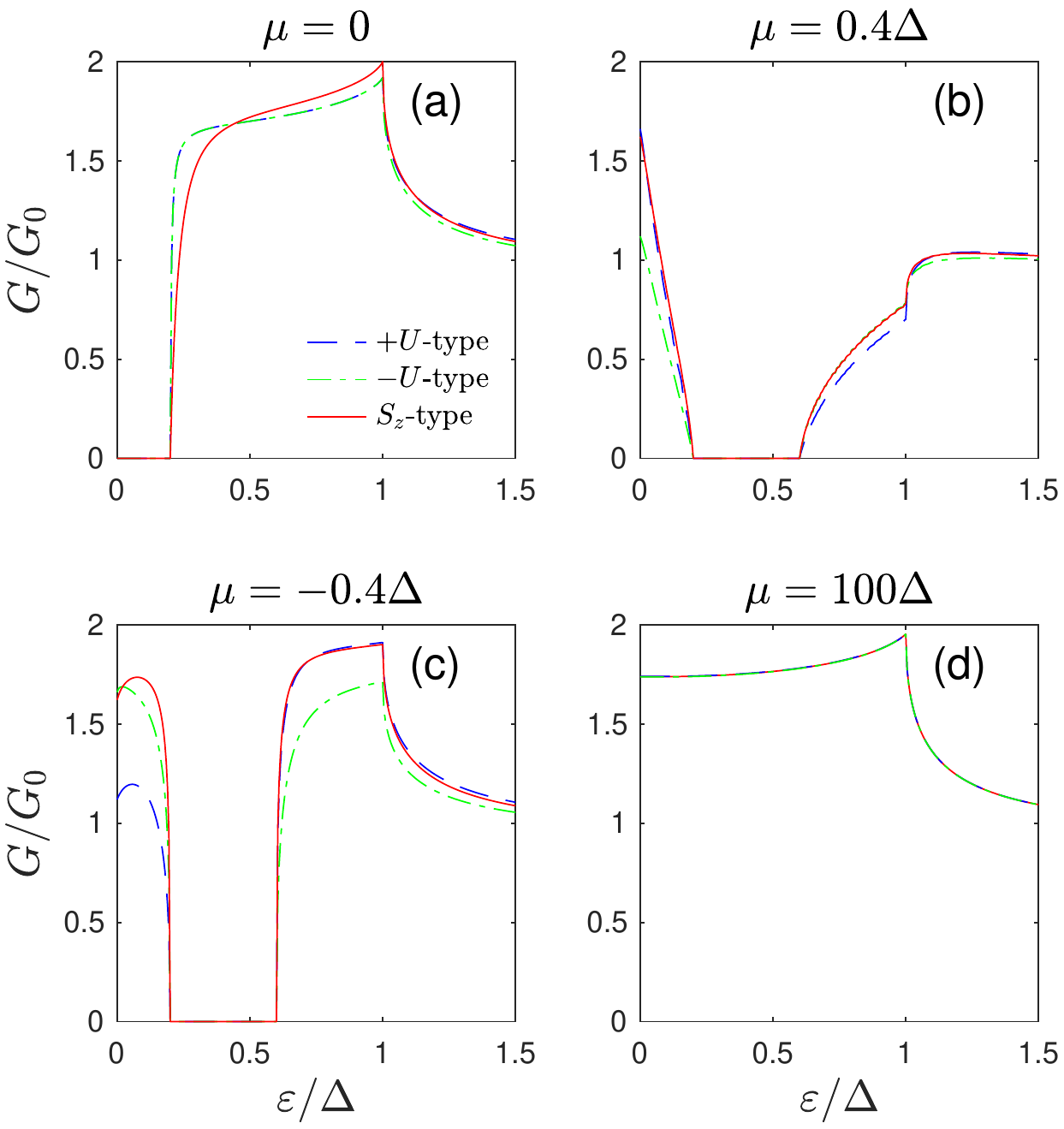}}
\caption{\label{f6}
Conductance spectrum with $m=0.2\Delta$.
}
\end{figure}

\subsection{Subgap conductance}
The subgap conductance $G/G_{0}$ is calculated from Eq.\ (\ref{GG}) with $m=0.2\Delta$ and plotted in Fig.\ \ref{f6}. In an undoped junction ($\mu =0$), the conductance is zero with $\varepsilon<m$ due to the lack of the propagating Andreev modes, as shown in Fig.\ \ref{f6}(a). After $\varepsilon=m$, the conductance increases with the incident energy due to the increasing of the density of states of the valence band holes. For the $S_z$-type fermions, the conductance reaches a peak $G/G_0=2$ at $\varepsilon=\Delta$ indicating the super Andreev reflection. The subgap conductances for the $\pm U$-type fermions are the same and also reach a peak at $\varepsilon=\Delta$, however, the peak value is less than 2 showing that the super Andreev reflection is attenuated.

In the $n$-doped junction ($\mu=0.4\Delta$), the Andreev reflection is in the retro-regime with the incident energy $\varepsilon <\mu-m$. The conductance of the $+U$-type fermions is larger than that of the $-U$-type fermions due to the oblique enhancement. The Andreev reflection is in the specular regime with the incident energy $\varepsilon >\mu+m$ and the oblique enhanced Andreev reflection occurs for the $-U$-type fermions so that the conductance of the $+U$-type fermions becomes smaller than that of the $-U$-type fermions as shown in Fig.\ \ref{f6}(b). In the interval $\mu-m<\varepsilon <\mu+m$, the Andreev process is prohibited leading to the zero conductance.

In the $p$-doped junction ($\mu=-0.4\Delta$), the conductance of the $+U$-type fermions is lager (smaller) than that of the $-U$-type fermions in the specular (retro-) regime, as shown in Fig.\ \ref{f6}(c), which is just the opposite result of that in the $n$-doped junction. 

In a heavily doped junction ($\mu=100\Delta$), the influence of the mass terms is negligible and the conductance of the different massive pseudospin-1 fermions are the same, as shown in Fig.\ \ref{f6}(d).

\section{Conclusions}\label{conclusions}

In conclusion, we have theoretically investigated the Andreev reflection of the massive pseudospin-1 fermions. It is found that the Andreev reflection for the $\pm U$-type fermions can be enhanced at the oblique incidence when the junction is doped ($\mu\neq0$). For the $+U$-type fermions the enhancement occurs in the retro-reflection (specular reflection) in the $n$-doped ($p$-doped) junction. For the $-U$-type fermions the enhancement occurs in the retro-reflection (specular reflection) in the $p$-doped ($n$-doped) junction. For the undoped junction ($\mu=0$), the oblique enhanced Andreev reflection is absent and the Andreev reflection probabilities for the $\pm U$-type fermions become identical. For the $S_z$-type fermions, there is a super Andreev reflection with unit efficiency independent of the incident angles when the incident energy approaches the superconducting gap. The super Andreev reflection is attenuated to a mass dependent quantity less than 1 for the $\pm U$-type fermions.
    
\section*{Acknowledgements}
This work is supported by the National Key R\&D Program of China (Grant No.\ 2017YFA0303203) and by the NSFC (Grant No.\ 11474149).

\end{document}